\documentclass[twocolumn,aps,pre,superscriptaddress,letterpaper]{revtex4-1}

\pdfoutput=1

\usepackage{graphicx}
\usepackage[caption=false]{subfig}
\usepackage{amsmath}
\usepackage{color}

\usepackage[utf8]{inputenc}
\definecolor{mpip} {rgb}{0,0.490196078431,0.478431372549}
\definecolor{mpip2} {rgb}{0,0.690196078431,0.278431372549}
\definecolor{grau} {rgb}{.5,.5,.5}
\definecolor{refokay} {rgb}{.1,.9,.7}

\begin{document}


\title{
Force-induced elastic matrix-mediated interactions in the presence of a rigid wall
}

\author{Andreas M. Menzel}
\email{menzel@thphy.uni-duesseldorf.de}
\affiliation{Institut f{\"u}r Theoretische Physik II: Weiche Materie, 
Heinrich-Heine-Universit{\"a}t D{\"u}sseldorf, D-40225 D{\"u}sseldorf, Germany}

\date{March 4, 2017}



\begin{abstract}
%
We consider an elastic composite material containing particulate inclusions in a soft elastic matrix that is bounded by a rigid wall, e.g., the substrate. If such a composite serves as a soft actuator, forces are imposed on or induced between the embedded particles. We investigate how the presence of the rigid wall affects the interactions between the inclusions in the elastic matrix. For no-slip boundary conditions, we transfer Blake's derivation of a corresponding Green's function from low-Reynolds-number hydrodynamics to the linearly elastic case. Results for no-slip and free-slip surface conditions are compared to each other and to the bulk behavior. Our results suggest that walls with free-slip surface conditions are preferred when they serve as substrates for soft actuators made from elastic composite materials. As we further demonstrate, the presence of a rigid wall can qualitatively change the interactions between the inclusions. In effect, it can switch attractive interactions into repulsive ones (and vice versa). It should be straightforward to observe the effects in future experiments and to combine our results, e.g., with the modeling of biological cells and tissue on rigid surfaces. 
\end{abstract}

\maketitle

\section{Introduction}

From low-Reynolds-number hydrodynamics \cite{purcell1977life,dhont1996introduction}, we know that the presence of a rigid wall can profoundly change the dynamic behavior of suspended objects. Frequently, no-slip boundary conditions are considered for the fluid on the surface. That is, at the positions of contact, the flow field of the suspending fluid vanishes. 
Then, for instance, hydrodynamic lift of polymers or vesicles away from the wall emerges when these objects move along the surface \cite{sendner2008shear,messlinger2009dynamical}. 
The dynamics of beating cilia is described correctly only by taking into account their hydrodynamic interaction with the anchoring substrate \cite{osterman2011finding,downton2009beating} and the effect of the wall in the framework of hydrodynamic synchronization has been worked out \cite{
wollin2011metachronal, golestanian2011hydrodynamic,uchida2012hydrodynamic}. 
Hydrodynamic interactions of self-propelling microswimmers with the surface can lead to effective attraction to and repulsion from the wall, depending on the propulsion mechanism \cite{elgeti2015physics,berke2008hydrodynamic}. 
Moreover, hydrodynamic coupling with the wall provides a breaking of symmetry that, for instance, allows a net forward motion of filaments composed of magnetic beads and rotated by an external magnetic field \cite{sing2010controlled}. 

Theoretically, in low-Reynolds-number hydrodynamics, the presence of such a wall is taken into account by a Green's function for the hydrodynamic Stokes equation that satisfies the boundary condition. By definition, it describes the fluid flow field induced by a point-like force center located within the fluid. It replaces the well-known Oseen tensor \cite{dhont1996introduction}, i.e., the Green's function for an infinitely extended bulk fluid. The direct derivation in the presence of a no-slip wall using Fourier transformations in the coordinates parallel to the surface dates back to a work by Blake \cite{blake1971note}, which is why the resulting Green's function is typically referred to as the Blake tensor. The resulting expression can be interpreted in an illustrative way. Similarly to the mirrored image charges in electrostatics \cite{jackson1962classical}, a system of mirror objects is placed behind the surface of the bounding wall. Their role is to ensure that the flow field vanishes on the surface of the wall. In effect, one may then again consider the problem for a virtually infinitely extended fluid, now additionally containing the mirror-image system.%

It is interesting to note that Blake in his original paper \cite{blake1971note} remarked a close connection to similar problems described by linear elasticity theory \cite{landau1986theory}. That is, to bodies that do not feature a terminal flow but instead deform reversibly according to a linearly elastic behavior. In fact, several of the methods derived in the hydrodynamic framework to characterize hydrodynamic suspensions have been transferred to the description of elastic bulk materials containing rigid inclusions \cite{phan1994load,kim1994faxen,norris2008faxen,puljiz2016forces,
puljiz2016long}. Particularly, this pertains to the description of deformation-mediated interactions via the embedding elastic matrix, the analogue of hydrodynamic interactions in the fluid case. More precisely, if forces are externally imposed on or induced between rigid inclusions in an elastic matrix, they lead to deformations of the matrix. These deformations are long-ranged and affect the positions of other inclusions, leading to matrix-mediated coupled displacements \cite{puljiz2016forces}. 

We here demonstrate that one can likewise obtain the Green's function for the linearly elastic problem in the presence of a rigid wall following Blake's direct calculation scheme familiar from hydrodynamics (Sec.~\ref{blake}). This further connects the two topical areas, low-Reynolds-number hydrodynamics and linear elasticity theory. The result is compared to a previous solution of the problem using a different approach \cite{phan1983image}. In contrast to the hydrodynamic situation, compressibility of the elastic matrix is readily included. 

We address the illustrative meaning of the resulting image systems 
(Sec.~\ref{image}). Then, we derive the framework to describe to lowest order the coupled displacement of several particulate inclusions in an elastic matrix near a rigid wall. These particles are subject to imposed forces, while they are additionally interacting with each other by inducing deformations of the embedding elastic matrix (Sec.~\ref{interactions}). For illustration, the results for two pairwisely interacting particles near a rigid wall are displayed. We consider the extreme situations of the parallel and perpendicular alignment of their connecting axis with the wall (Sec.~\ref{results}). Interestingly, we find that the presence of the wall cannot only quantitatively influence but in effect can even \textit{reverse the resulting relative displacements} of the particles.  
That is, the wall may in effect \textit{reverse attraction and repulsion} between the inclusions (Sec.~\ref{reversal}). 
It should be possible to observe this effect in future experiments.  

Our results should be interesting for the characterization of elastic composite materials consisting of rigid inclusions embedded in an elastic environment. One prospective application of such materials is their use as soft actuators \cite{an2003actuating,filipcsei2007magnetic, fuhrer2009crosslinking,bose2012soft}. For instance, external magnetic or electric fields may induce interactions between the inclusions and lead to overall distortions \cite{diguet2009dipolar,stolbov2011modelling,wood2011modeling, ivaneyko2012effects,zubarev2013effect,menzel2015tuned, allahyarov2015simulation,huang2016buckling,metsch2016numerical}, or net forces are imposed onto the inclusions when they are drawn into an external magnetic field gradient \cite{zrinyi1996deformation}. Our situation corresponds to the contact area where the composite material is placed on a suitable substrate. In corresponding experiments, also free-slip boundary conditions could be realized, using, for example, a lubricant on the surface of the substrate. 
We therefore obtain our results for these modified boundary conditions as well. In conclusion (Sec.~\ref{conclusion}), our results suggest an advantage in enabling free-slip surface conditions on the substrates in actuator applications.

\section{Derivation of the Green's functions}
\label{blake}

Distorted states of elastic bodies in linear elasticity theory \cite{landau1986theory} are described by a displacement field $\mathbf{u}(\mathbf{x})$ that describes the reversible relocations of the volume elements from their initial positions. We first derive explicit expressions for the displacement fields caused by a point-like force center. 
In other words, we derive an explicit expression of the corresponding elastic Green's functions. 
Our derivation for no-slip boundary conditions follows the same scheme as the one presented by Blake in Ref.~\onlinecite{blake1971note} and thus demonstrates the close connection between low-Reynolds-number hydrodynamics and the linearly elastic case. 

For this purpose, we consider a semi-infinitely extended homogeneous isotropic elastic matrix. The matrix is bounded by a no-slip surface located at $z=0$, i.e., $\mathbf{u}(z=0)=\mathbf{0}$. We confine ourselves to small-amplitude deformations so that linear elasticity theory applies. 
In (quasi-)static situations, linearly elastic behavior is governed by the Navier-Cauchy equation \cite{cauchy1828exercises},
\begin{equation}\label{navier-cauchy}
	\nabla^2\mathbf{u}(\mathbf{x}) + \frac{1}{1-2\nu}\nabla\nabla\cdot\mathbf{u}(\mathbf{x}) ={} -\frac{1}{\mu}\,
	\mathbf{F}\delta(\mathbf{x}-\mathbf{x}_0).
\end{equation}
Here, $\mu$ is the shear modulus of the elastic matrix, $\nu$ is its Poisson ratio connected to the matrix compressibility, $\delta(\bullet)$ denotes the delta function, and $\mathbf{F}$ represents a point force acting onto the matrix at position $\mathbf{x}_0$. This basic equation plays the same role as the Stokes equation \cite{dhont1996introduction} for the derivation of the Blake tensor in low-Reynolds-number hydrodynamics \cite{blake1971note}. There, the derivation is restricted to incompressible systems, and an additional equation that sets the divergence of the flow field to zero needs to be satisfied. In our case, this additional condition is absent and compressibility of the matrix is readily included. 

For an infinitely extended bulk elastic matrix, the solution to the Navier-Cauchy equation is given by $\mathbf{u}(\mathbf{x})=\mathbf{\hspace{.02cm}\underline{\hspace{-.02cm}G}}(\mathbf{x}-\mathbf{x}_0) \cdot\mathbf{F}$, where the Green's function $\mathbf{\hspace{.02cm}\underline{\hspace{-.02cm}G}}(\mathbf{r})$ is obtained via Fourier transformation methods and is given by \cite{landau1986theory,weinberger2005,puljiz2016long}
\begin{equation}\label{greens_function}
	\mathbf{\hspace{.02cm}\underline{\hspace{-.02cm}G}}(\mathbf{r}) ={} \frac{1}{16\pi(1-\nu)\mu}\left[\frac{3-4\nu}{r}\mathbf{\underline{\hat{I}}}+\frac{\mathbf{r}\mathbf{r}}{r^3}\right].
\end{equation}
We mark second-rank tensors and matrices by an underscore, denote the unity matrix by $\mathbf{\underline{\hat{I}}}$, by $\mathbf{r}\mathbf{r}$ the dyadic product, and set $r=\|\mathbf{r}\|$. 

\begin{figure}
\includegraphics[width=8.3cm]{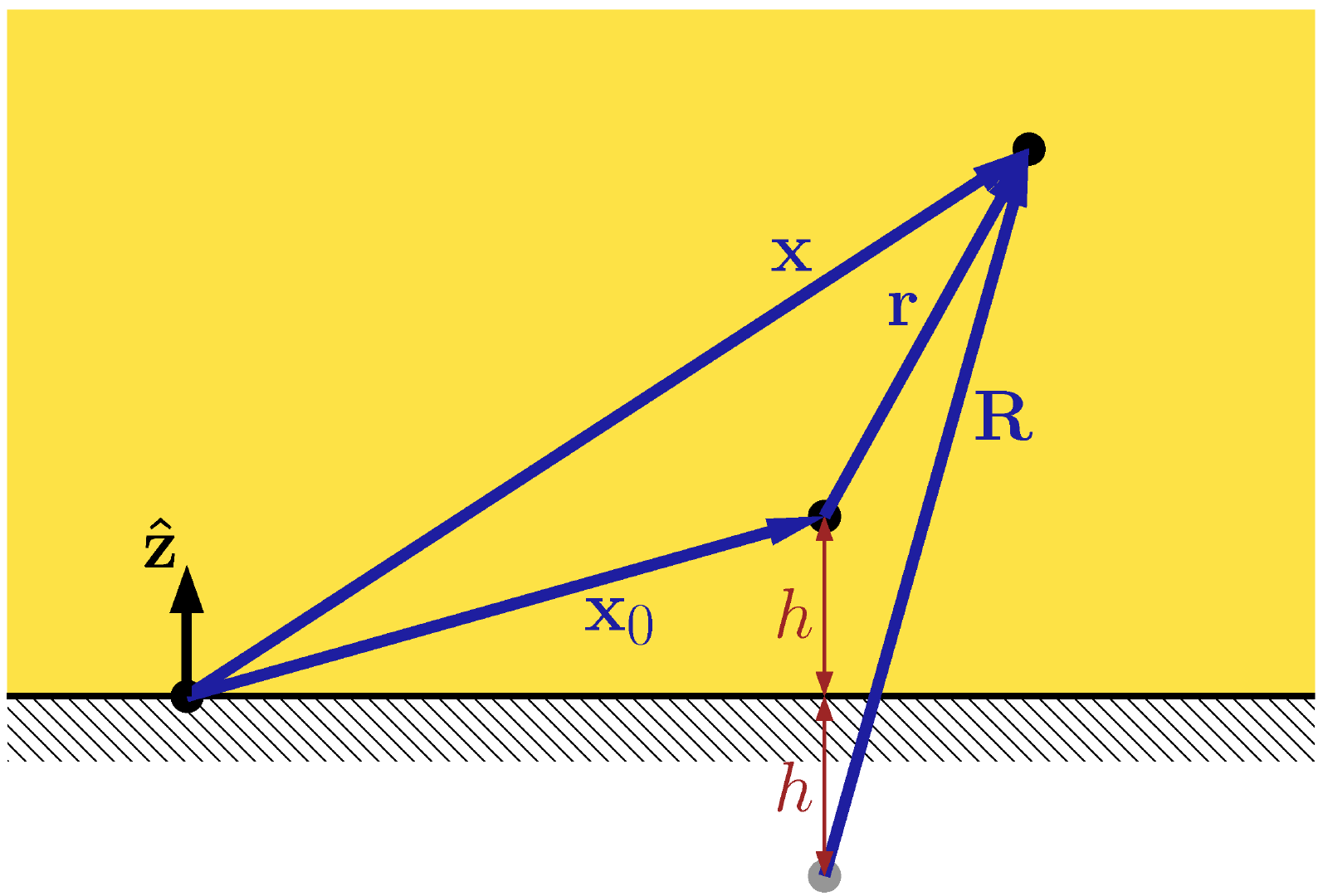}
\caption{Schematic illustration of the geometry to determine the image system of the Green's function. The half-space $z>0$ is filled by an elastic matrix with no-slip boundary conditions at $z=0$. A point force is applied at a position $\mathbf{x}_0$ of height $z=h$ above the boundary. We then search for a mirror-image system at $z=-h$ to ensure the no-slip boundary condition. The position of a given spot $\mathbf{x}$ within the matrix relatively to the location of the point force and its image is denoted by $\mathbf{r}$ and $\mathbf{R}$, respectively.}
\label{fig_geometry}
\end{figure}

Our point force is located at a ``height'' (distance) $h>0$ above the rigid wall, i.e., at $z=h$, see also Fig.~\ref{fig_geometry}. 
The reasoning then is the same as, for instance, in the electrostatics example of a point charge located in front of a planar conducting wall \cite{jackson1962classical}. We apply the Green's function in Eq.~(\ref{greens_function}) as if the whole space were filled by an infinitely extended elastic matrix. As a consequence, displacements result on the no-slip surface at $z=0$. Then we search for an image system at $z=-h$, behind the surface, that counteracts the displacements at $z=0$ and ensures the no-slip boundary condition $\mathbf{u}(z=0)=\mathbf{0}$. 

We follow Blake's notation in that the location of a given spot $\mathbf{x}$ within the matrix relatively to the location $\mathbf{x}_0$ of the point force is denoted by $\mathbf{r}=\mathbf{x}-\mathbf{x}_0$ and relatively to the image system by $\mathbf{R}=\mathbf{x}-\mathbf{x}_0+2h\mathbf{\hat{z}}$, see Fig.~\ref{fig_geometry}. Thus, $\mathbf{R}=\mathbf{r}+2h\mathbf{\hat{z}}$. We call our searched-for Green's function $\mathbf{\hspace{.02cm}\underline{\hspace{-.02cm}B}}(\mathbf{r})$ and, using Eq.~(\ref{greens_function}), choose the ansatz \cite{blake1971note}
\begin{equation}\label{ansatz}
\mathbf{\underline{B}}(\mathbf{r}) =
\mathbf{\hspace{.02cm}\underline{\hspace{-.02cm}G}}(\mathbf{r})
- \mathbf{\hspace{.02cm}\underline{\hspace{-.02cm}G}}(\mathbf{R})
+ \mathbf{\hspace{.02cm}\underline{\hspace{-.02cm}W}}(\mathbf{R}). 
\end{equation}
That is, an oppositely oriented point force has been placed at the mirrored position at $z=-h$. The central task is then to determine the remaining part $\mathbf{\hspace{.02cm}\underline{\hspace{-.02cm}W}}(\mathbf{R})$ from the remaining equation
\begin{equation}\label{eq_W}
\nabla^2\mathbf{\hspace{.02cm}\underline{\hspace{-.02cm}W}}
+\frac{1}{1-2\nu}\nabla\nabla\cdot\mathbf{\hspace{.02cm}\underline{\hspace{-.02cm}W}}
=
\mathbf{\underline{0}}
\end{equation}
obtained upon insertion of $\mathbf{u}(\mathbf{x})=\mathbf{\underline{B}}(\mathbf{r})\cdot\mathbf{F}$ into Eq.~(\ref{navier-cauchy}). 

At this point, it appears beneficial to introduce an auxiliary variable 
\begin{equation}\label{def_P}
\mathbf{P}
=\nabla\cdot\mathbf{\hspace{.02cm}\underline{\hspace{-.02cm}W}}.
\end{equation}
It takes a role analogous to the pressure function in the hydrodynamic case \cite{blake1971note}. In the end, we need to enforce that Eq.~(\ref{def_P}) be satisfied. Taking the divergence of Eq.~(\ref{eq_W}), we further find
\begin{equation}\label{P_harmonic}
\nabla^2 \mathbf{P}
=\mathbf{0}.
\end{equation}

The boundary condition for $\mathbf{\hspace{.02cm}\underline{\hspace{-.02cm}W}}(\mathbf{R})$ on the no-slip surface at $z=0$ follows from Eqs.~(\ref{greens_function}) and (\ref{ansatz}) by requiring $\mathbf{\underline{B}}(\mathbf{r}|_{z=0})=\mathbf{\underline{0}}$. At $z=0$, we have $r|_{z=0}=R|_{z=0}=:R_0$. Using the notation of Ref.~\onlinecite{blake1971note}, we denote by Greek indices only the in-plane coordinates, numbered by $1$ and $2$, while $3$ marks the perpendicular $z$-coordinate. Roman indices run through all three coordinates. As a result, we obtain for the no-slip boundary condition
\begin{eqnarray}\label{RB}
W_{ij}(\mathbf{R}|_{z=0}) &=& W_{ij}(R_1,R_2,h) \nonumber\\
&=& \frac{h}{8\pi(1-\nu)\mu}\frac{R_{\alpha}}{R_0^3}\left(\delta_{i3}\delta_{j\alpha}+\delta_{i\alpha}\delta_{j3}\right), 
\end{eqnarray}
where $\delta_{\bullet\bullet}$ denotes the Kronecker delta and summation over repeated indices is implied. 

As in the hydrodynamic case \cite{blake1971note}, we then perform a Fourier transform in the in-plane coordinates 
\begin{equation}\label{FT}
{\mathcal{F\!\!T}}\{\bullet\}=\frac{1}{2\pi}\int_{-\infty}^{\infty}\mathrm{d}R_1\int_{-\infty}^{\infty}\mathrm{d}R_2\;\bullet\; \mathrm{e}^{\mathrm{i}(k_1R_1+k_2R_2)}. 
\end{equation}
Abbreviating $k=\sqrt{k_1^2+k_2^2}$, we obtain from Eqs.~(\ref{eq_W}) and (\ref{P_harmonic})
\begin{eqnarray}
0 &=& {}-k^2\tilde{W}_{ij} +\frac{\partial^2}{\partial R_3^2}\tilde{W}_{ij}
\nonumber\\ 
&&\quad{}+ \frac{1}{1-2\nu}\left(-\mathrm{i}\,k_{\alpha}\delta_{i\alpha}+\delta_{i3}\frac{\partial}{\partial R_3}\right)\tilde{P}_j, \\
0 &=& {}-k^2\tilde{P}_j + \frac{\partial^2}{\partial R_3^2}\tilde{P}_j,
\end{eqnarray}
where the tilde marks the Fourier-transformed quantities. 

The solutions of these equations read
\begin{eqnarray}
\tilde{P}_j &=& B_j\,\mathrm{e}^{-kR_3}, \\
\tilde{W}_{ij} &=& B_{ij}\,\mathrm{e}^{-kR_3} - \frac{1}{2(1-2\nu)}\left(\mathrm{i}\,\frac{k_{\alpha}}{k}\delta_{i\alpha}+\delta_{i3}\right)B_j\nonumber\\
&&{}\qquad\qquad\qquad\qquad\quad\times(R_3-h)\,\mathrm{e}^{-kR_3}.
\label{FT_W}
\end{eqnarray}
Here, the coefficients $B_j$ and $B_{ij}$ are constant with respect to $R_3$. We obtain $B_{ij}$ by satisfying the no-slip boundary condition at $z=0$. For this purpose, we determine the Fourier transform of Eq.~(\ref{RB}) by applying Eq.~(\ref{FT}). To perform the double integral, we found it convenient to switch to polar coordinates. As a result, we obtain
\begin{equation}
\tilde{W}_{ij}(k_1,k_2,h) = \frac{h}{8\pi(1-\nu)\mu}\
\mathrm{i}\left(\delta_{i3}\delta_{j\alpha}+\delta_{i\alpha}\delta_{j3}\right) 
\frac{k_{\alpha}}{k}\mathrm{e}^{-kh}.
\end{equation}
Comparing with Eq.~(\ref{FT_W}) for $R_3|_{z=0}=h$ leads to 
\begin{equation}\label{eqBij}
B_{ij} = \frac{h}{8\pi(1-\nu)\mu}\
\mathrm{i}\left(\delta_{i3}\delta_{j\alpha}+\delta_{i\alpha}\delta_{j3}\right) 
\frac{k_{\alpha}}{k}.
\end{equation}
After that, we find $B_j$ by enforcing that the Fourier transform of Eq.~(\ref{def_P}), i.e.,
\begin{equation}
\tilde{P}_j = \left(-\mathrm{i}\,k_{\alpha}\delta_{i\alpha}+\delta_{i3}\,\frac{\partial}{\partial R_3}\right)\tilde{W}_{ij}, 
\end{equation}
be satisfied, resulting in 
\begin{equation}\label{eqBj}
B_j = \frac{2(1-2\nu)}{3-4\nu}\left(-\mathrm{i}\,k_{\alpha}\delta_{i\alpha}-k\,\delta_{i3}\right)B_{ij}.
\end{equation}
Combining Eqs.~(\ref{FT_W}), (\ref{eqBij}), and (\ref{eqBj}), we have derived the Fourier-transformed components $\tilde{W}_{ij}$. 

\begin{widetext}
Next, it is a straightforward calculation to perform on $\tilde{W}_{ij}$ the Fourier transform inverse to Eq.~(\ref{FT}). Inserting the result into Eq.~(\ref{ansatz}), we obtain
\begin{eqnarray}
B_{ij}(\mathbf{r}) &=& 
\frac{1}{16\pi(1-\nu)\mu}
\left[(3-4\nu)\left(\frac{1}{r}-\frac{1}{R}\right)\delta_{ij}+\frac{r_ir_j}{r^3}-\frac{R_iR_j}{R^3}\right] 
+\frac{h}{8\pi(1-\nu)\mu}\,\frac{1}{R^3}
\Bigg[
\left(\delta_{i\alpha}\delta_{j3}+\delta_{i3}\delta_{j\alpha}\right)R_{\alpha} 
\nonumber
\end{eqnarray}
\begin{eqnarray}
{}\quad\qquad\qquad-\frac{R_3-h}{3-4\nu}\:
\Bigg\{
\delta_{i\alpha}\delta_{j\beta}\left(\delta_{\alpha\beta}-3\,\frac{R_{\alpha}R_{\beta}}{R^2}\right)
+3\left(\delta_{i\alpha}\delta_{j3}-\delta_{i3}\delta_{j\alpha}\right)\frac{R_{\alpha}R_3}{R^2}
-\delta_{i3}\delta_{j3}\left(1-3\,\frac{R_3^2}{R^2}\right)
\Bigg\}
\Bigg].
\end{eqnarray}
This expression, with $\alpha,\beta\in\{1,2\}$, matches Blake's hydrodynamic result, if we identify the shear modulus $\mu$ with the hydrodynamic viscosity and set $\nu=0.5$ for an incompressible system (to enable the direct comparison, we need to explicitly carry out the derivative in Eq.~(16) of Ref.~\onlinecite{blake1971note}). 

Finally, we switch back to exclusively Roman indices, which leads us to
\begin{eqnarray}
B_{ij}(\mathbf{r}) &=& 
\frac{1}{16\pi(1-\nu)\mu}
\left[(3-4\nu)\left(\frac{1}{r}-\frac{1}{R}\right)\delta_{ij}+\frac{r_ir_j}{r^3}-\frac{R_iR_j}{R^3}\right] 
\nonumber\\[.1cm]
&&
{}+\frac{h}{8\pi(1-\nu)\mu}\,\frac{1}{R^3}
\bigg[
\delta_{i3}R_j+\delta_{j3}R_i-2\,\delta_{i3}\delta_{j3}R_3 
+\frac{R_3-h}{3-4\nu}\:
\bigg(
2\,\delta_{i3}\delta_{j3}-\delta_{ij}+3\,\frac{R_iR_j}{R^2}-6\,\frac{R_iR_3}{R^2}\delta_{j3}
\bigg)
\bigg].
\label{B}
\quad
\end{eqnarray}
\end{widetext}
The latter result has been obtained before by a different method within the context of linear elasticity theory in Ref.~\onlinecite{phan1983image} (where, however, in the result of Ref.~\onlinecite{phan1983image} we find that a factor of $R^3$ should be deleted from the denominator in the second line of Eq.~(29); moreover, we think that the shear modulus is missing in three denominators of Eq.~(25) and a minus sign should be added to the first expression of Eq.~(26) in Ref.~\onlinecite{phan1983image}). 

We here showed that Blake's direct approach using the in-plane Fourier transform, which is the common approach for related problems in hydrodynamic systems \cite{blake1971note,liron1976stokes}, can be transferred to the linearly elastic case as well. This should help to further connect these two subfields of classical continuum mechanics in the future. An auxiliary variable in analogy to the hydrodynamic pressure field has been introduced intermediately to facilitate the calculation. 

For our later comparison we here further introduce the Green's function $\mathbf{\underline{C}}(\mathbf{r})$ for a free-slip boundary at $z=0$. By this, we understand a surface satisfying the boundary condition $u_z(z=0)=0$. That is, the matrix may freely slip along the rigid wall parallel to the surface, but it may not penetrate into or detach from the wall. 
This condition is met if we mirror at $z=0$ any force acting within the matrix, see also the hydrodynamic case \cite{mathijssen2015tracer}. That is, the in-plane coordinates of the force are maintained, but the normal coordinate is inverted. We may express this inversion by defining a modified bulk Green's function $\mathbf{\hspace{.02cm}\underline{\hspace{-.02cm}G}}^m(\mathbf{R})$, the components of which reading
\begin{equation}\label{Gm}
G^m_{ij}(\mathbf{R})={}-\left(j^2-3j+1\right)\,G_{ij}(\mathbf{R}).
\end{equation}
In the brackets, the function in $j$ ensures that the inversion due to the leading minus sign only becomes effective in the normal coordinate, i.e., for $j=3$. 
Then, the components of the resulting Green's function $\mathbf{\underline{C}}(\mathbf{r})$ that correctly contains the mirror image may be denoted as
\begin{equation}\label{Green_fs}
C_{ij}(\mathbf{r})=G_{ij}(\mathbf{r})+G_{ij}^m(\mathbf{R}).
\end{equation}
It is straightforward to verify that $C_{ij}(\mathbf{r})$ vanishes for $i=3$ as required.

\section{Illustration of the image system}
\label{image}

From the construction of the Green's function $\mathbf{\underline{C}}(\mathbf{r})$, the nature of the associated mirror-image system of a point-like force center close to a rigid free-slip surface is obvious. It is given by mirroring at the plane $z=0$ the initial point force, see Eq.~(\ref{navier-cauchy}), 
as expressed by $\mathbf{\hspace{.02cm}\underline{\hspace{-.02cm}G}}^m(\mathbf{R})$ in Eq.~(\ref{Green_fs}). But how does the mirror-image system in the case of a no-slip surface look like? 

This mirror-image system must be represented by the Green's function $\mathbf{\underline{B}}(\mathbf{r})$ in Eq.~(\ref{B}), except for the terms in $r$ that represent the initial point force and correspond to the bulk Green's function $\mathbf{\hspace{.02cm}\underline{\hspace{-.02cm}G}}(\mathbf{r})$. All other terms solely depend on $\mathbf{R}$. Thus the mirror-image system is completely located at the position of the initial point force mirrored at the plane $z=0$. The remaining part of the first square bracket in Eq.~(\ref{B}) corresponds to $-\mathbf{\hspace{.02cm}\underline{\hspace{-.02cm}G}}(\mathbf{R})$ and thus represents the inverted initial point force located at the mirrored position. 
It turns out that the terms between the second square brackets in Eq.~(\ref{B}) can be interpreted similarly to the hydrodynamic case \cite{blake1971note}, plus modifications and additional effects due to the possible compressibility. 

\begin{figure}
\includegraphics[width=8.3cm]{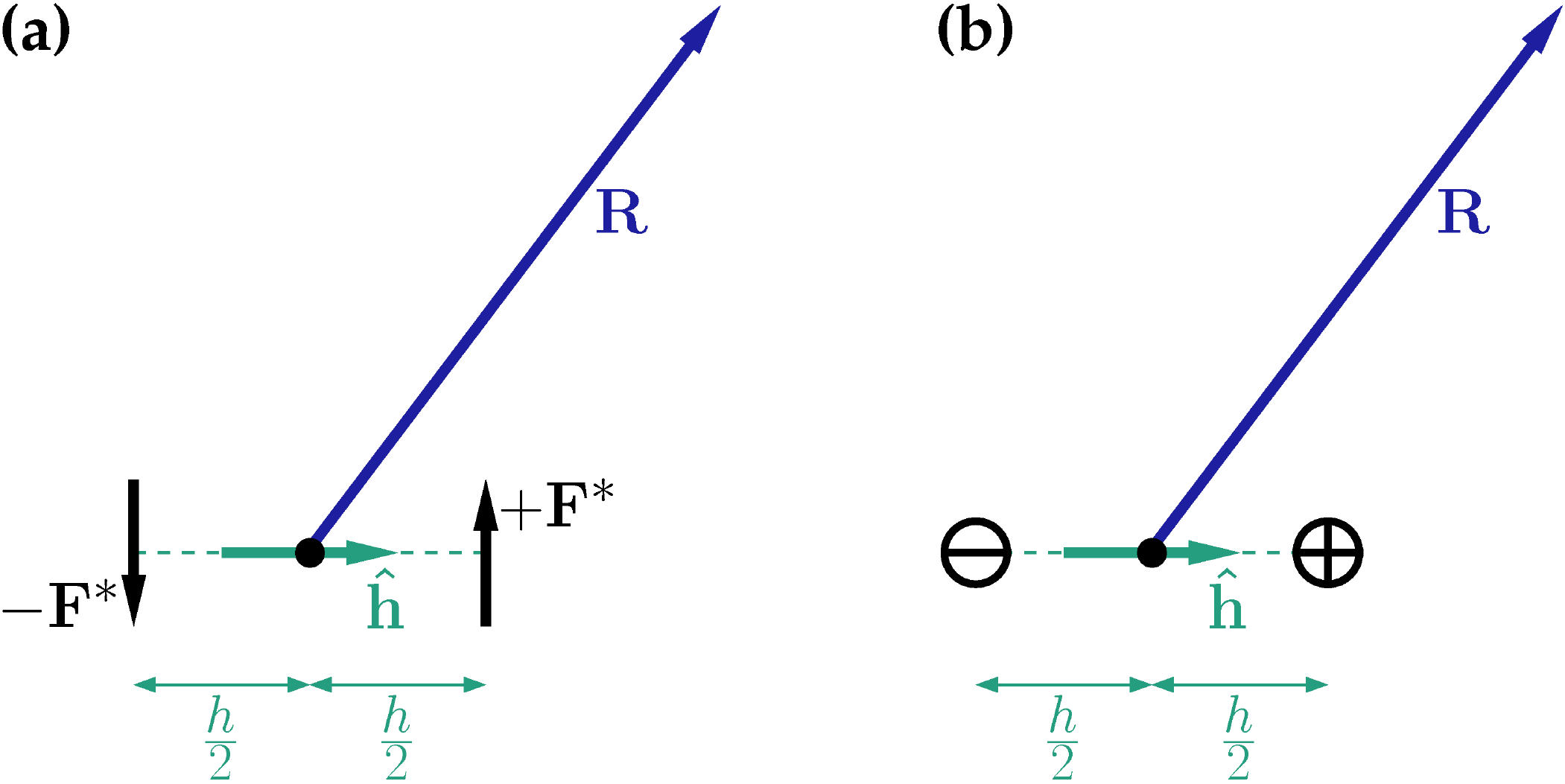}
\caption{(a) Set-up of a force doublet. The point of observation is located a distance $\mathbf{R}$ away from the center between the two forces. From that center, it is a distance $+\frac{h}{2}\mathbf{\hat{h}}$ to the force $+\mathbf{F}^*$ and a distance $-\frac{h}{2}\mathbf{\hat{h}}$ to the force $-\mathbf{F}^*$. (b) Analogous set-up for a source doublet. Here, the source $\oplus$ and the sink $\ominus$ are located at $+\frac{h}{2}\mathbf{\hat{h}}$ and at $-\frac{h}{2}\mathbf{\hat{h}}$ from the center of the doublet, respectively.}
\label{fig_doublet}
\end{figure}

To make progress, we first calculate the displacement field $\mathbf{u}^{fd}(\mathbf{R})$ induced by a combination of two antiparallel forces $\pm\mathbf{F}^*$, the point-like force centers of which being separated by a distance vector $\mathbf{h}=h\mathbf{\hat{h}}$, see Fig.~\ref{fig_doublet}(a). 
The displacement field induced by this force doublet follows via the Green's function in Eq.~(\ref{greens_function}). We expand it to first order in $h$:
\begin{eqnarray}
u_i^{fd}(\mathbf{R}) 
&=& 
G_{ij}\left(\mathbf{R}-\frac{h}{2}\mathbf{\hat{h}}\right)F_j^*
- G_{ij}\left(\mathbf{R}+\frac{h}{2}\mathbf{\hat{h}}\right)F_j^* 
\nonumber\\[.1cm]
&\approx&
\frac{h}{8\pi(1-\nu)\mu}\,\frac{1}{R^3}\bigg[
R_l\hat{h}_l\delta_{ij}+\frac{1}{3-4\nu}\bigg(
-R_i\hat{h}_j\quad
\nonumber\\[.1cm]
&&
\qquad{}-R_j\hat{h}_i+3\frac{R_iR_jR_l\hat{h}_l}{R^2}\bigg)
\bigg]
\frac{3-4\nu}{2}F_j^*.
\label{fd}
\end{eqnarray}
We compare this expression with the displacement field $\mathbf{u}(\mathbf{r})=\mathbf{\underline{B}}(\mathbf{r})\cdot\mathbf{F}$ caused by a point force $\mathbf{F}$ and calculated via the Green's function in Eq.~(\ref{B}). 

\begin{figure}
\includegraphics[width=8.3cm]{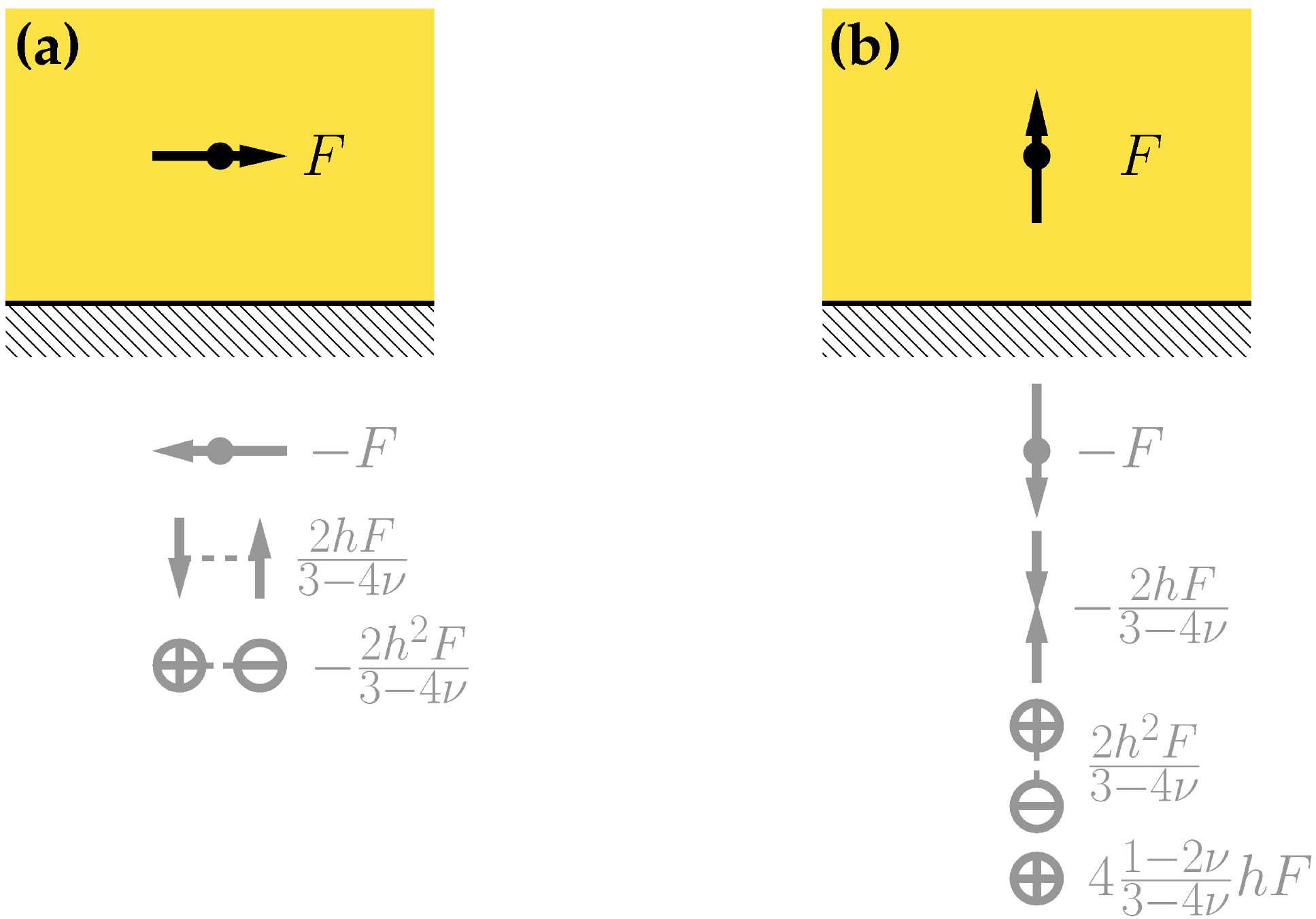}
\caption{Illustration of the mirror-image systems. (a) For a point force $\mathbf{F}$ oriented parallel to the no-slip surface, the image system consists of an oppositely oriented point force, a force doublet, and a source doublet of the indicated strengths. (b) If $\mathbf{F}$ points away from (towards) the no-slip surface, the image system consists of an oppositely oriented point force, a force doublet, a source doublet, and an additional source (sink) of the given strengths. In both cases, inversion of the initial point force $\mathbf{F}$ also inverts all forces of the mirror-image systems and simultaneously swaps sources and sinks.}
\label{fig_imagesystem}
\end{figure}

First, if $\mathbf{F}$ is oriented parallel to the no-slip surface, the contribution due to the second square bracket in Eq.~(\ref{B}) reduces to 
\begin{equation}
\frac{h}{8\pi(1-\nu)\mu}\,\frac{1}{R^3}
\left[
\delta_{i3}R_{\alpha} 
+\frac{R_3}{3-4\nu}\:
\left(
-\delta_{i\alpha}+3\,\frac{R_iR_{\alpha}}{R^2}
\right)
\right]F_{\alpha},
\end{equation}
where $\alpha\in\{1,2\}$ and the terms $\sim h^2$ will be addressed separately below. 
We notice that this expression coincides with the one in Eq.~(\ref{fd}), if we choose $\mathbf{\hat{h}}\|\mathbf{F}$, i.e., $\mathbf{F}=F\mathbf{\hat{h}}$, and $F^*_j = 2F/(3-4\nu)\delta_{j3}$. In other words, this part of the mirror-image system in effect represents a force doublet with the forces perpendicular and their connecting vector parallel to the no-slip surface, see also Fig.~\ref{fig_imagesystem}(a). 
Defining the strength $M_{fd}$ of this force doublet as the magnitude of the forces times the separation distance, we obtain 
\begin{equation}
M_{fd} = \frac{2hF}{3-4\nu}.
\end{equation}
In the incompressible limit, i.e., for $\nu=0.5$, this value reproduces the one derived by Blake \cite{blake1971note}. We note at this point that any separation distance smaller than $h$ could have been used for the derivation as well, if at the same time we increase the strength of the effective force $\mathbf{F}^*$ accordingly. This remark allows the same picture even if $h$ itself does not represent a small expansion parameter. 

The situation changes, if $\mathbf{F}$ is oriented perpendicular to the no-slip surface, i.e., $\mathbf{F}=F\mathbf{\hat{z}}$. Then, we obtain from
the second square bracket in Eq.~(\ref{B})
\begin{eqnarray}
\lefteqn{\frac{h}{8\pi(1-\nu)\mu}\,\frac{1}{R^3}
\bigg[
-\delta_{i3}R_3 +R_i}
\nonumber\\
&&
{}\qquad\qquad\qquad+\frac{R_3}{3-4\nu}\:
\bigg(
\delta_{i3}-3\,\frac{R_iR_3}{R^2}
\bigg)
\bigg]F_3,\quad
\label{fd_perp}
\end{eqnarray}
where again the terms $\sim h^2$ will be addressed separately below. 
In this case, we find almost complete agreement with Eq.~(\ref{fd}), if we set $\mathbf{\hat{h}}=\mathbf{\hat{z}}$ and $F^*_j=-2F/(3-4\nu)\delta_{j3}$. This choice corresponds to a force doublet of strength 
\begin{equation}
M_{fd} = {}-\frac{2hF}{3-4\nu},
\end{equation}
see Fig.~\ref{fig_imagesystem}(b). 
However, when we compare Eqs.~(\ref{fd}) and (\ref{fd_perp}), we notice a slight difference associated with the first appearance of $R_i$. This difference amounts to an additional contribution 
\begin{equation}
\frac{h}{8\pi(1-\nu)\mu}\;2\,\frac{1-2\nu}{3-4\nu}\,\frac{R_i}{R^3}\;F
\end{equation}
in the image system beyond the pure force doublet. Apparently, this extra part vanishes in the incompressible case for $\nu=0.5$. It corresponds to an isolated point source or sink, and we refer to it as a sourcelet. Here, we define a sourcelet via its induced displacement field 
\begin{equation}\label{s}
\mathbf{u}^s(\mathbf{R}) = \frac{1}{16\pi(1-\nu)\mu}\,\frac{\mathbf{R}}{R^3}\,M_s,
\end{equation}
where $M_s$ is its strength. This displacement field satisfies the homogeneous part of the Navier-Cauchy equation Eq.~(\ref{navier-cauchy}), which is readily seen by recasting $\mathbf{R}/R^3={}-\nabla\, 1/R$. Obviously, in our case
\begin{equation}
M_s = 4\,\frac{1-2\nu}{3-4\nu}\,hF. 
\end{equation}
That is, we have an additional mirror-image source behind the no-slip surface, if $\mathbf{F}$ points away from the wall ($F>0$). Vice versa, we find a mirror-image sink, if $\mathbf{F}$ points towards the wall ($F<0$). There are dilations and compressions of the matrix between the point force and the wall for a compressible system. For $\nu=0.5$, when the elastic matrix is incompressible, these contributions vanish, and an image system analogous to the hydrodynamic case is observed \cite{blake1971note}. 

Finally, we address the illustrative background of the remaining terms $\sim h^2$ in Eq.~(\ref{B}). For this purpose, we first determine the displacement field $\mathbf{u}^{sd}(\mathbf{R})$ induced by a combination of one source and one sink separated by a distance vector $\mathbf{h}=h\mathbf{\hat{h}}$, see Fig.~\ref{fig_doublet}(b). Proceeding in analogy to our above treatment for the force doublet, we here use Eq.~(\ref{s}) to calculate the resulting displacement field for this source doublet. Expanding to first order in $h$, we obtain
\begin{eqnarray}
u_i^{sd}(\mathbf{R}) 
&=& 
u^s_i\left(\mathbf{R}-\frac{h}{2}\mathbf{\hat{h}}\right)
- u^s_i\left(\mathbf{R}+\frac{h}{2}\mathbf{\hat{h}}\right) 
\nonumber\\[.1cm]
&\approx&
\frac{h}{16\pi(1-\nu)\mu}\,\frac{1}{R^3}\left(
{}-\hat{h}_i+3\,\frac{R_iR_l\hat{h}_l}{R^2}
\right)
M_{s}^*,
\nonumber\\
&&
\label{sd}
\end{eqnarray}
where $M^*_s$ is the strength of the underlying effective sources.

First, we consider again a force $\mathbf{F}$ applied parallel to the no-slip plane. Then the terms $\sim h^2$ in Eq.~(\ref{B}) reduce to 
\begin{equation}
\frac{h}{8\pi(1-\nu)\mu}\,\frac{1}{R^3}
\left[
\frac{-h}{3-4\nu}\:
\left(
{}-\delta_{i\alpha}+3\,\frac{R_iR_{\alpha}}{R^2}
\right)
\right]F_{\alpha},
\end{equation}
where $\alpha\in\{1,2\}$. This expression coincides with Eq.~(\ref{sd}), if we choose $\mathbf{\hat{h}}\|\mathbf{F}$, i.e., $\mathbf{F}=F\mathbf{\hat{h}}$, and $M^*_s={}-2hF/(3-4\nu)$. Thus, this part of the mirror-image system represents a source doublet with the separation vector between both sourcelets parallel to the no-slip surface, see Fig.~\ref{fig_imagesystem}(a). The strength of this source doublet $D_{sd}=hM^*_{s}$ is therefore given by
\begin{equation}\label{Dsdpar}
D_{sd} = {}-\frac{2h^2F}{3-4\nu}. 
\end{equation}

Second, for $\mathbf{F}$ oriented perpendicular to the no-slip surface, i.e., $\mathbf{F}=F\mathbf{\hat{z}}$, we find for the terms $\sim h^2$ in Eq.~(\ref{B})
\begin{equation}
\frac{h}{8\pi(1-\nu)\mu}\,\frac{1}{R^3}
\left[
\frac{-h}{3-4\nu}\:
\left(
{}\delta_{i3}-3\,\frac{R_iR_3}{R^2}
\right)
\right]F_3.
\end{equation}
Agreement with Eq.~(\ref{sd}) is achieved by setting $\mathbf{\hat{h}}=\mathbf{\hat{z}}$ and $M_s^*=2hF/(3-4\nu)$. Thus, in this case, the separation vector between the two sourcelets is oriented perpendicular to the no-slip boundary, see Fig.~\ref{fig_imagesystem}(b). The strength of the effective source doublet follows as
\begin{equation}\label{Dsdperp}
D_{sd} = \frac{2h^2F}{3-4\nu}.
\end{equation}
Both, Eqs.~(\ref{Dsdpar}) and (\ref{Dsdperp}), reproduce their hydrodynamic counterparts derived by Blake \cite{blake1971note} when considering the incompressible limit for $\nu=0.5$. We have here described the illustrative meaning of the complete mirror-image system for compressible elastic matrices.

\section{Inclusion interactions}
\label{interactions}

Our central concern is to demonstrate how displacements within the matrix that are coupled via the induced matrix deformations are affected by the presence of the rigid wall. 
We here think of small particulate inclusions in the elastic matrix. Forces on these particles can be imposed from outside or induced between them, e.g., by external electric or magnetic fields or field gradients. 
As in related hydrodynamic approaches \cite{wollin2011metachronal,dufresne2000hydrodynamic,squires2000like}, we only consider the leading-order matrix-mediated couplings between the particles, i.e., couplings to first order in the inverse separation distance between the particles. To this order, our previous comparison with experimental measurements in the bulk confirmed a very good match 
for moderate particle separation distances \cite{puljiz2016forces}. 

If a force $\mathbf{F}_j$ is acting on the \textit{r}eal particle $j$, this force is transmitted to the embedding surrounding matrix. In the presence of a no-slip surface, our Green's function in Eq.~(\ref{ansatz}) gives the resulting displacement field at any position $\mathbf{x}$ within the matrix, $\mathbf{u}_j^{(r)}(\mathbf{x})=\mathbf{\underline{B}}(\mathbf{x}-\mathbf{x}_j)\cdot\mathbf{F}_j$. Since the other particles are embedded and anchored in the matrix, they are displaced together with the field induced by particle $j$. This leads to a displacement of the $i$th particle as given by the matrix displacement field at position $\mathbf{x}_i$, i.e., 
\begin{equation}\label{U1r}
\mathbf{U}_i^{(1,r)}=\mathbf{u}_j^{(r)}(\mathbf{x}_i)=
\mathbf{\underline{B}}(\mathbf{r}_{ij})\cdot\mathbf{F}_j, 
\end{equation}
where $\mathbf{r}_{ij}=\mathbf{x}_i-\mathbf{x}_j$. 

Moreover, if particle $i$ is direct subject to a force $\mathbf{F}_i$, an additional direct displacement $\mathbf{U}_i^{(0)}$ results. Here, we recall the notion behind the mirror-image approach. In effect, we treat the matrix as infinitely extended, filling the whole space. Yet, a superimposed mirror-image system is placed at the mirror position behind the no-slip surface to satisfy the no-slip boundary condition. 

Therefore, to lowest order, the direct displacement of a spherical particle $i$ of radius $a$ is given by the bulk expression for an isolated particle \cite{phan1993rigid,phan1994load,kim1994faxen, puljiz2016forces,puljiz2016long}, 
\begin{equation}\label{M0}
\mathbf{U}_i^{(0)}=M_0\,\mathbf{F}_i, \quad  M_0 =\frac{5-6\nu}{24\pi(1-\nu)\mu a},
\end{equation}
which represents the elastic analogue to the hydrodynamic Stokes solution \cite{dhont1996introduction}. However, particle $i$ simultaneously interacts with its own \textit{m}irror-image system. The mirror-image system leads to additional matrix displacements that we read off from Eq.~(\ref{ansatz}) as $\mathbf{u}_i^{(m)}(\mathbf{x})=\left[- \,\mathbf{\hspace{.02cm}\underline{\hspace{-.02cm}G}}(\mathbf{R})
+ \mathbf{\hspace{.02cm}\underline{\hspace{-.02cm}W}}(\mathbf{R})\right]\cdot\mathbf{F}_i$. Thus, particle $i$ itself is additionally displaced as prescribed by the displacement field induced by its own mirror-image system. At the particle position $\mathbf{x}_i$, this leads to the additional displacement 
\begin{equation}\label{U1m}
\mathbf{U}^{(1,m)}_i=\mathbf{u}_i^{(m)}(\mathbf{x}_i)=\left[{}- \mathbf{\hspace{.02cm}\underline{\hspace{-.02cm}G}}(\mathbf{R}_i)
+ \mathbf{\hspace{.02cm}\underline{\hspace{-.02cm}W}}(\mathbf{R}_i)\right]\cdot\mathbf{F}_i,
\end{equation} 
where $\mathbf{R}_i=2x_{i,3}\mathbf{\hat{z}}$. 

Due to the linearity of the underlying Eq.~(\ref{navier-cauchy}), we may simply superimpose the different contributions from Eqs.~(\ref{U1r})--(\ref{U1m}). Moreover, we may add the influence of further particles in an analogous way. In total, the coupled displacements of $N$ particles are given by
\begin{equation}\label{displaceability}
\left(
\begin{array}{c}
\mathbf{U}_1 \\ \vdots \\[.1cm] \mathbf{U}_N
\end{array}
\right)
=
\left(
\begin{array}{ccc}
\mathbf{\underline{M}}_{11}
& \cdots & \mathbf{\underline{M}}_{1N}
\\ 
\vdots & \ddots & \vdots \\[.1cm]
\mathbf{\underline{M}}_{N1}
& \cdots & \mathbf{\underline{M}}_{NN}
\end{array}
\right)
\cdot
\left(
\begin{array}{c}
\mathbf{F}_1 \\ \vdots \\[.1cm] \mathbf{F}_N
\end{array}
\right).
\end{equation}
Here, $\mathbf{\underline{M}}_{ij}
$ ($i,j=1,...,N$) are the \textit{displaceability matrices}. In the presence of a rigid no-slip boundary, they read
\begin{eqnarray}
\mathbf{\underline{M}}_{i=j}^{ns} &=& M_0\,\mathbf{\underline{\hat{I}}} - \mathbf{\hspace{.02cm}\underline{\hspace{-.02cm}G}}(\mathbf{R}_i)
+ \mathbf{\hspace{.02cm}\underline{\hspace{-.02cm}W}}(\mathbf{R}_i),
\label{Mnsii}
\\[.1cm]
\mathbf{\underline{M}}_{i\neq j}^{ns} &=& \mathbf{\underline{B}}(\mathbf{r}_{ij}),
\label{Mnsij}
\end{eqnarray}
where we had defined $\mathbf{r}_{ij}=\mathbf{x}_i-\mathbf{x}_j$ and $\mathbf{R}_i=2x_{i,3}\mathbf{\hat{z}}$, the components of $\mathbf{\underline{B}}(\mathbf{r})$ are given by Eq.~(\ref{B}), while the components of $\left[-\, \mathbf{\hspace{.02cm}\underline{\hspace{-.02cm}G}}(\mathbf{R})
+ \mathbf{\hspace{.02cm}\underline{\hspace{-.02cm}W}}(\mathbf{R})\right]$ follow from Eq.~(\ref{B}) by omitting the two terms containing $r$. 

The appealing character of Eq.~(\ref{displaceability}) is that the role of the elastic matrix is implicitly contained in the displaceability matrices. If we know the positions of the inclusions, we can directly calculate these matrices via Eqs.~(\ref{Mnsii}) and (\ref{Mnsij}). We do not need to explicitly resolve the distortions of the elastic matrix itself. If we further know the forces on all particles, their coupled displacements result from Eq.~(\ref{displaceability}) via simple matrix multiplication. 

In the case of a free-slip surface, Eq.~(\ref{displaceability}) formally applies in the same way. Using the corresponding Green's function defined in Eqs.~(\ref{Gm}) and (\ref{Green_fs}), the displaceability matrices then read
\begin{eqnarray}
\mathbf{\underline{M}}_{i=j}^{fs} &=& M_0\,\mathbf{\underline{\hat{I}}} + \mathbf{\hspace{.02cm}\underline{\hspace{-.02cm}G}}^m(\mathbf{R}_i),
\label{Mfsii}
\\[.1cm]
\mathbf{\underline{M}}_{i\neq j}^{fs} &=& \mathbf{\underline{C}}(\mathbf{r}_{ij}).
\label{Mfsij}
\end{eqnarray}
In the bulk, we simply have
\begin{eqnarray}
\mathbf{\underline{M}}_{i=j}^{b} &=& M_0\,\mathbf{\underline{\hat{I}}},
\label{Mbii}
\\[.1cm]
\mathbf{\underline{M}}_{i\neq j}^{b} &=& \mathbf{\hspace{.02cm}\underline{\hspace{-.02cm}G}}(\mathbf{r}_{ij}).
\label{Mbij}
\end{eqnarray}

In order to illustrate the effect of the rigid wall, we must evaluate these expressions. For this purpose, we specify the forces acting on the particles. 
For instance, we have recently investigated the behavior of magnetizable paramagnetic Nickel particles embedded in a bulk elastic matrix \cite{puljiz2016forces}. There, magnetic interactions  between the particles were induced and tuned by applying and rotating an external magnetic field. 
For identical spherical magnetizable particles of radius $a$ in a saturating homogeneous external magnetic field, the induced magnetic dipole moment $\mathbf{m}=m\mathbf{\hat{m}}$ ($m=\|\mathbf{m}\|$) is identical for all particles. It scales as $m\sim a^3$. For two particles $i$ and $j$ located at positions $\mathbf{x}_i$ and $\mathbf{x}_j$, respectively, the induced pairwise dipolar interaction force is given by \cite{jackson1962classical}
\begin{equation}\label{magneticforce}
\mathbf{F}_i=-\mathbf{F}_j=
-\frac{3\mu_0m^2\!\left[5\mathbf{\hat{r}}_{ij}(\mathbf{\hat{m}}\!\cdot\!\mathbf{\hat{r}}_{ij})^2-\mathbf{\hat{r}}_{ij}-2\mathbf{\hat{m}}(\mathbf{\hat{m}}\!\cdot\!\mathbf{\hat{r}}_{ij})\right]}{4\pi\,r_{ij}^4},
\end{equation}
with $\mu_0$ the magnetic vacuum permeability, $\mathbf{r}_{ij}=\mathbf{x}_i-\mathbf{x}_j$, $r_{ij}=\|\mathbf{r}_{ij}\|$, and $\mathbf{\hat{r}}_{ij}=\mathbf{r}_{ij}/r_{ij}$ ($i\neq j$). Obviously, this force changes with relative displacements of the particles. In the calculations below, we include these corrections by a simple iterative loop to determine the final displacements and forces \cite{puljiz2016forces}. 

In summary, we work here with pairwise magnetic forces. This underlines, for instance, the importance of our results for magnetic elastic composite materials serving as soft actuators \cite{filipcsei2007magnetic,fuhrer2009crosslinking,bose2012soft} and the transferability of such approaches to corresponding electric situations \cite{allahyarov2015simulation,allahyarov2016dipole}. However, our considerations naturally apply in the same way for any other, not necessarily pairwise forces acting on the particles in an elastic environment.

\section{Impact of the rigid wall}
\label{results}

For illustration, we confine ourselves to a pair of embedded magnetizable particles. We measure all lengths in units of $a$. Moreover, we set the magnitude of the induced saturated volume magnetization of our magnetic particles to $M=30\sqrt{\mu/\mu_0}$, where $m=(4\pi/3)a^3M$. In SI units, these numbers correspond, for instance, to $\mu=100~\mathrm{Pa}$ and $M=267~\mathrm{kA/m}$, which are close to the ones inferred for our previous experimental investigation in a bulk system \cite{puljiz2016forces}. Accordingly, we also choose an initial separation between the particles of distance $d=7a$. Then, presenting all displacements in units of the particle radius $a$, our results are independent of the exact particle size. 

\subsection{Parallel configuration}\label{sec_par}

\begin{figure}
\includegraphics[width=8.3cm]{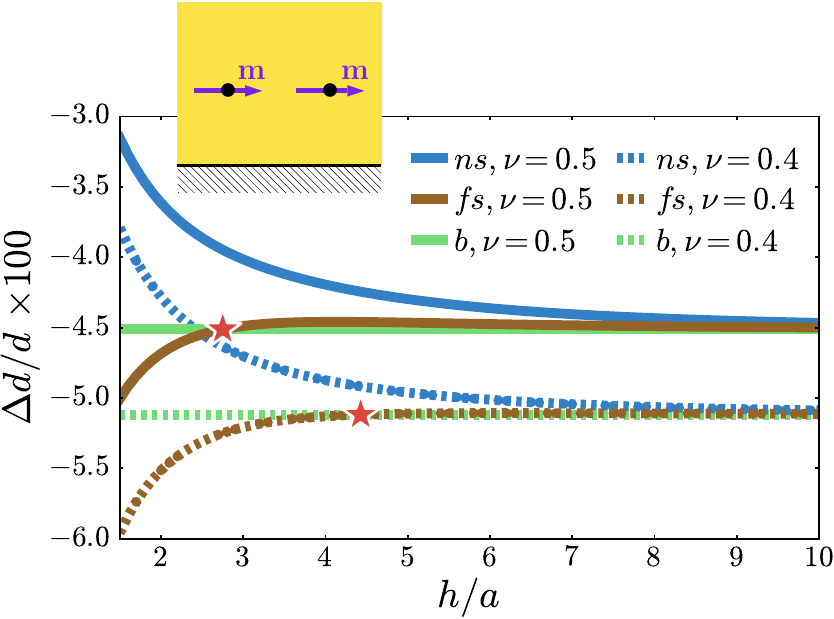}
\caption{Two particles in parallel configuration next to a rigid wall are magnetized such that attractive magnetic interactions between the particles arise ($M=30\sqrt{\mu/\mu_0}$). We compare the results for no-slip ($ns$) and free-slip ($fs$) boundary conditions to those in the bulk ($b$) of the matrix. The relative change in distance $\Delta d/d$ between the particles is plotted as a function of the ``height'' $h$ above the surface, here for an initial distance $d=7a$, with $a$ the particle radius. A hindering effect of the no-slip surface is obvious. Free-slip surfaces support the particle approach at lower height $h$ and, for Poisson ratios $\nu>0.25$, slightly counteract at larger height $h$. Solid lines are for $\nu=0.5$, dotted lines for $\nu=0.4$. The stars mark the crossing points between the results for the free-slip boundary and the bulk.}
\label{fig_parallel}
\end{figure}

We start with a configuration of two particles having their separation vector oriented parallel to the rigid wall, see Fig.~\ref{fig_parallel}. 
An attractive magnetic force between the particles is applied by setting $\mathbf{\hat{m}}=\mathbf{\hat{x}}$. Then, we calculate the resulting displacements in the presence of a no-slip boundary, a free-slip surface, and in the bulk according to Eqs.~(\ref{displaceability})--(\ref{magneticforce}). The relative change in distance $\Delta d/d$ is plotted in Fig.~\ref{fig_parallel} as a function of the height $h$ above the surface. Here, we observe an approach of the particles due to their mutual attraction. In the repulsive case, we obtain analogous results with the particles displacing away from each other. 

When comparing with the bulk values, given by the horizontal lines in Fig.~\ref{fig_parallel}, the influence of the rigid wall is obvious. For a no-slip boundary, we observe a significantly reduced change in distance close to the surface. Particularly, the mirror-image forces are oriented oppositely to the real forces, see Fig.~\ref{fig_imagesystem}(a), and impede the displacements of the corresponding real-side particles. 

In contrast to that, we observe a stronger approach for free-slip boundary conditions. At lower heights $h$, this approach is even stronger than in the bulk. The reason is that in the free-slip case, the particles during their approach do not have to drag along the whole matrix of the lower half-space. 
Thinking in terms of mirror-image forces, the mirrored free-slip forces support the displacement of the particles on the real side, in contrast to the no-slip case, where they are oppositely oriented. 

However, at larger heights $h$, we observe that the free-slip change in distance is slightly reduced when compared to the bulk situation. From inspection of Eqs.~(\ref{Mfsii}) and (\ref{Mfsij}) in opposition to Eqs.~(\ref{Mbii}) and (\ref{Mbij}), we infer that it is the mirror-image force of the other particle that opposes to the displacement of each particle. Following this argument, if only one single particle is relocated due to a force applied parallel to the free-slip surface, we should always observe a higher displacement than in the bulk. We have checked that this is indeed the case. In some sense, a bit counterintuitively, we may therefore attribute the reduced approach of the two particles in the free-slip situation to the increased displaceability for each individual particle on its own. 

Moreover, from this inspection of Eqs.~(\ref{Mfsii})--(\ref{Mbij}), we find an expression for the height $h$ at which the relative changes in distance between the particles near the free-slip-boundary and in the bulk are equal. It is given by 
\begin{eqnarray}
h &=& \frac{d}{2\sqrt{2(4\nu-1)}}\Bigg[ \frac{32\nu^2-40\nu+11}{3-4\nu} 
\nonumber\\
&&
{}+\bigg( 
\frac{(32\nu^2-40\nu+11)^2}{(3-4\nu)^2}  
+4(3-4\nu)(4\nu-1)
\bigg)^{\!\!\frac{1}{2}}
\:
\Bigg]^{\!\frac{1}{2}}
\nonumber\\[.1cm]
&&
\end{eqnarray} 
and marked in Fig.~\ref{fig_parallel} by the stars.
This height diverges at the Poisson ratio $\nu=0.25$. For values $\nu<0.25$, there is no crossing between the free-slip and bulk values any longer. Then, the free-slip condition at all heights leads to higher relative approaches between the two particles than in the bulk situation. 

\begin{figure}
\includegraphics[width=8.3cm]{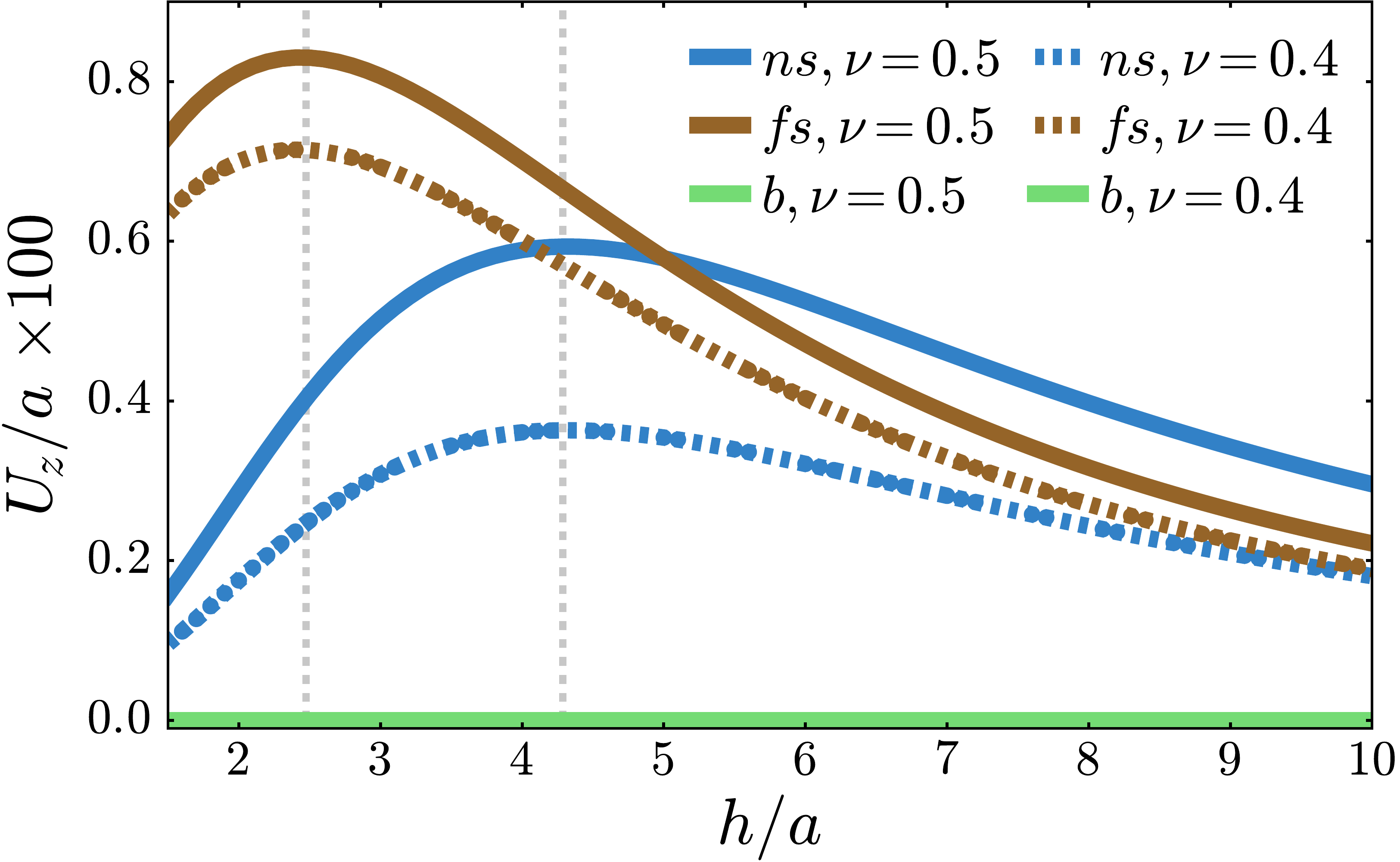}
\caption{Displacement $U_z$ in the direction away from the wall for the two particles considered in Fig.~\ref{fig_parallel}. A maximum occurs both for free-slip and no-slip boundary conditions as marked by the vertical dotted lines. The maximum lift away from the surface is higher in the free-slip case and for the incompressible system ($\nu=0.5$). In bulk, no displacement perpendicular to the connecting line between the particles occurs.}
\label{fig_uz_parallel}
\end{figure}

Apart from that, we observe in Fig.~\ref{fig_parallel} a higher magnitude of the particle approach when we turn from an incompressible matrix of $\nu=0.5$ to the compressible matrix of $\nu=0.4$. Thus the compressibility here supports the effect. Moreover, we see that the relative displacements are always larger for the free-slip wall than close to a no-slip surface. When we think of the application of such materials as soft actuators \cite{an2003actuating,filipcsei2007magnetic, fuhrer2009crosslinking,bose2012soft}, it is this relative change in distance between the particles that should typically be maximized. We therefore conclude that a free-slip surface would be a significantly more supportive choice of substrate for such a device when compared to a no-slip boundary. 

Furthermore, the interaction with the wall leads to a perpendicular relocation during the particle approach. We depict the corresponding displacements in Fig.~\ref{fig_uz_parallel}. 
\begin{figure}
\includegraphics[width=8.3cm]{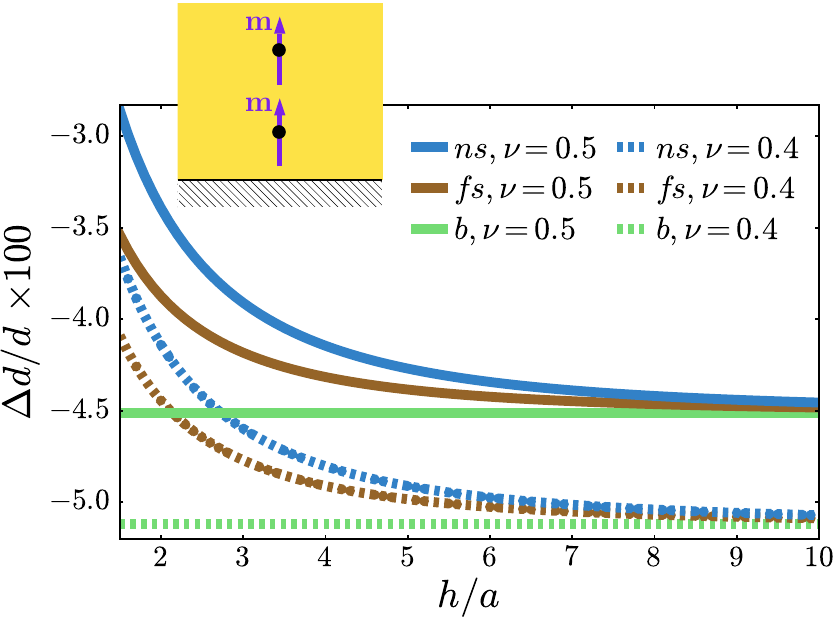}
\caption{Two particles in normal configuration next to a rigid wall, subject to attractive magnetic interactions between the particles. Again, the results for no-slip and free-slip boundary conditions are compared to those in the bulk of the matrix. Here, the relative change in distance $\Delta d/d$ between the particles is plotted as a function of the ``height'' $h$ of the lower particle above the surface. Remaining parameters are the same as in Fig.~\ref{fig_parallel}. Both surface conditions hinder the mutual approach, while the counteraction of the no-slip wall is stronger. Increasing the compressibility from the incompressible case of $\nu=0.5$ (solid lines) to a Poisson ratio of $\nu=0.4$ (dotted lines) allows for larger particle approaches.}
\label{fig_perp}
\end{figure}
When the inclusions due to their mutual attraction approach each other, they squeeze out matrix material from between them. Partially, this material is pressed towards the wall. This leads to an effective lift of the particles away from the surface. Mathematically, the lift follows for each particle from the mirror-image system of the other particle, see Fig.~\ref{fig_imagesystem}(a). In the no-slip case, all contributions are involved, i.e., the mirrored force, the force doublet, and the source doublet. We can calculate the magnitude of the lift from Eqs.~(\ref{Mnsii})--(\ref{Mfsij}) via the implicit mirror-image systems. As marked in Fig.~\ref{fig_uz_parallel} by the vertical dotted lines, a maximum lift occurs for free-slip conditions at $h=d/2\sqrt{2}$ and for no-slip conditions at $h=\sqrt{3}d/2\sqrt{2}$. The free-slip maximum value is above the no-slip one, thus, again, a larger particle displacement is possible for a free-slip surface. Moreover, we here observe that incompressibility ($\nu=0.5$) supports the particle relocation. Incompressible material squeezed out from between the particles is more effectively pressed towards the wall. 

\subsection{Normal configuration}

In the second step, we consider the axis connecting the two particles to be oriented perpendicular to the rigid wall, see Fig.~\ref{fig_perp}.
Again we consider attraction between the two particles, now by setting $\mathbf{\hat{m}}=\mathbf{\hat{z}}$. Results for repulsion follow accordingly. 

Fig.~\ref{fig_perp} shows the relative changes in distance $\Delta d/d$. Here, the height $h$ refers to the distance of the closer particle from the surface. As we can see, the presence of the rigid wall significantly reduces the mutual approach between the particles when compared to the bulk value. In contrast to the parallel configuration, also the free-slip surface is observed to hinder the induced displacements at all values of $h$. 

\begin{figure}
\includegraphics[width=8.3cm]{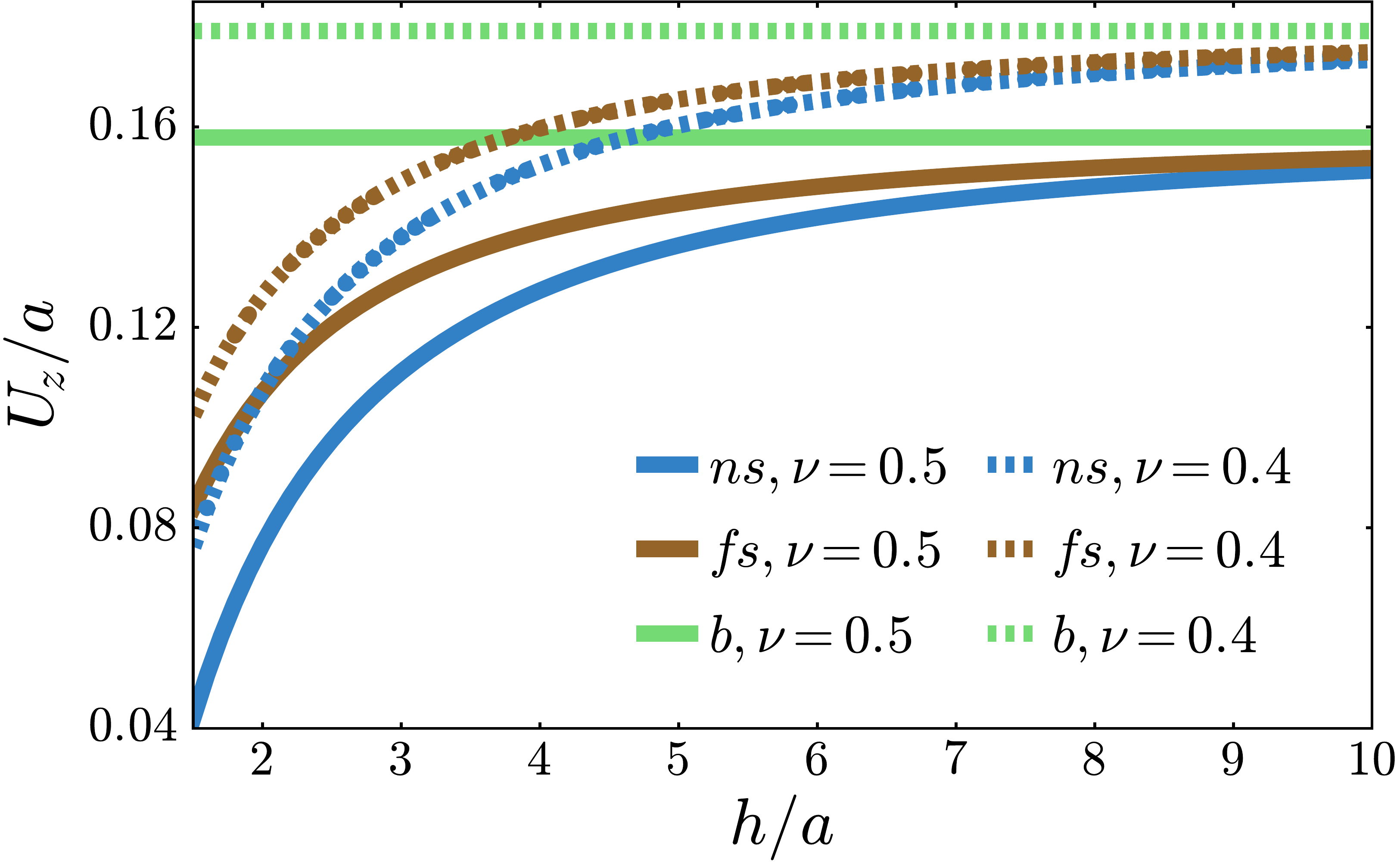}
\caption{Displacement in the direction away from the substrate for the particle closer to the wall in the set-up of Fig.~\ref{fig_perp}. Concentrating on this displacement to a big extent explains the behavior observed in Fig.~\ref{fig_perp}.}
\label{fig_uz_perp}
\end{figure}

To explain the behavior, we concentrate on the particle closer to the surface. It displaces away from the wall towards the other particle. We display its normal relocation in Fig.~\ref{fig_uz_perp}. 
During this relocation, the particle needs to take part of the surrounding matrix material along with it. However, the normal motion of matrix material is hindered by the wall. Thus, for compensation, more material needs to be pulled in from the sides. These lateral matrix displacements are additionally hindered on the no-slip surface. Increasing the distance $h$ from the wall, the bulk values are approached. Moreover, we again find in Figs.~\ref{fig_perp} and \ref{fig_uz_perp} that increasing the compressibility allows for larger displacements. 

In terms of the mirror-image systems, we obtain for both surface conditions, no-slip and free-slip, oppositely oriented mirror-image forces below the wall. Due to the smaller distances, particularly the particle closer to the wall is partially drawn towards the surface by its own mirror-image force. In the no-slip case, the additional force doublet further pulls the lower particle towards the wall and thus additionally reduces the magnitudes of displacement. For compressible elastic matrices, a supportive mirror-image source is introduced beyond the no-slip wall, see Fig.~\ref{fig_imagesystem}(b). 

Altogether, we find that the free-slip surface allows for stronger relative changes in distance than the no-slip wall, also in the normal configuration. These results support the free-slip wall as a candidate for a substrate in actuator applications. 

\section{Effective reversal of particle attraction and repulsion}
\label{reversal}

Finally, we come to an interesting and at first glance possibly unexpected \textit{inverting} effect of the rigid boundary in certain situations. Namely, due to the presence of the wall, induced \textit{attractive} interactions between embedded particles may in effect appear \textit{repulsive}, and vice versa. 
For elastically embedded magnetizable particles, it seems most practical to concentrate on the conversion of induced attraction into effective repulsion, although in theory the inverse situation could be described as well. 

\begin{figure}
\includegraphics[width=8.cm]{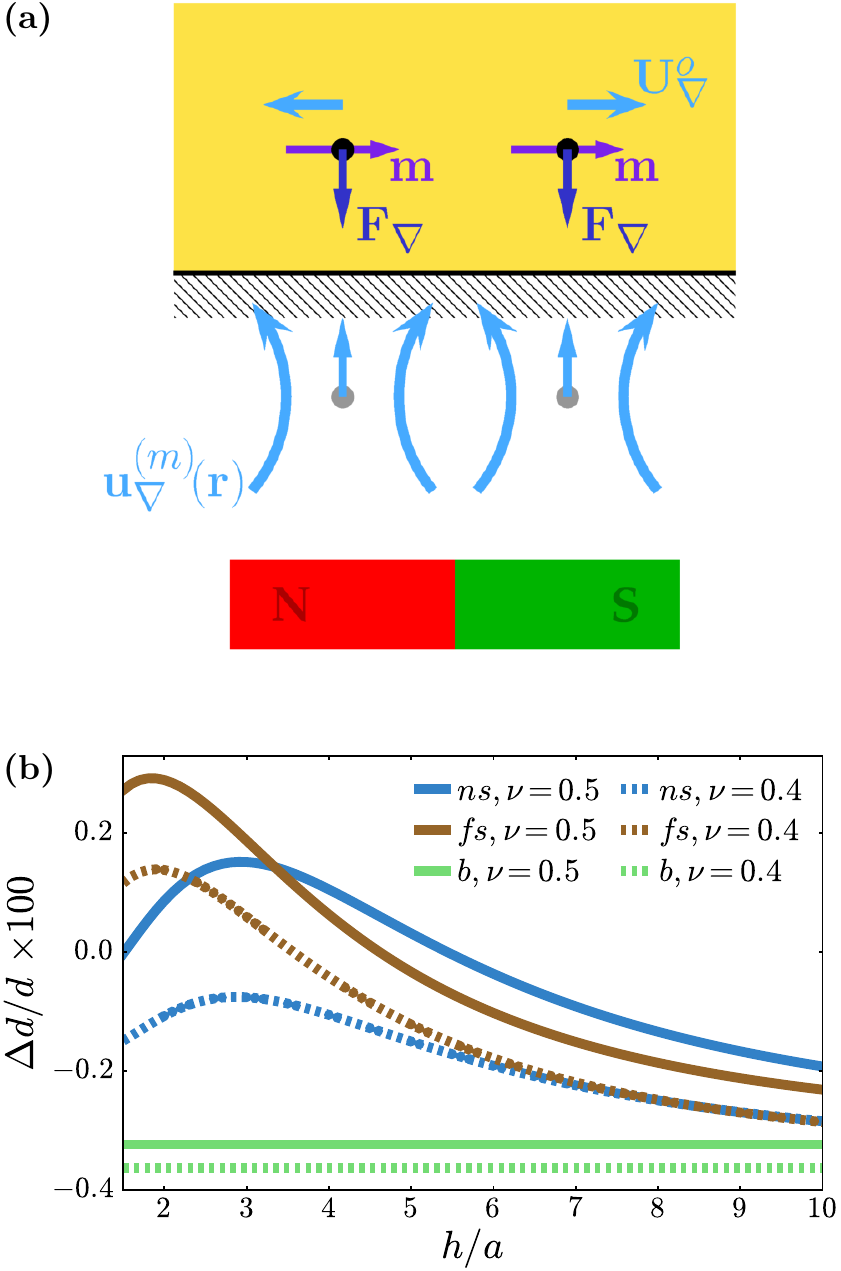}
\caption{Two magnetizable particles in parallel configuration above a rigid wall attracting each other, see Fig.~\ref{fig_parallel}, but here additionally pulled towards the surface. (a) This situation could, for instance, be realized by placing a permanent magnet underneath the substrate. Then, gradient forces arise that are oriented normal to the boundary. Their corresponding mirror-image forces point upward. The displacement fields $\mathbf{u}^{(m)}_{\nabla}(\mathbf{r})$ generated by the mirrored forces act on both real particles, leading to the outward displacements $\mathbf{U}_{\nabla}^o$. (b) Plotting the relative change in distance $\Delta d/d$ between the particles as a function of their height $h$ above the substrate demonstrates that the induced attraction between the particles can be effectively converted into repulsion by the wall ($\Delta d/d>0$). An incompressible elastic matrix ($\nu=0.5$, solid lines) is more supportive to the effect than the depicted compressible elastic matrix ($\nu=0.4$, dotted lines), and we observe higher maximal magnitudes for the free-slip boundary when compared to the no-slip boundary. In bulk the particles attract each other as expected ($\Delta d/d<0$). Here, we used a magnetic field gradient of $\partial_zB=-0.3\sqrt{\mu\mu_0}/a$, a particle magnetization $M=4\sqrt{\mu/\mu_0}$, and an initial distance of $d=5a$.}
\label{fig_reversal}
\end{figure}

Therefore, let us consider the parallel configuration of two particles as in Sec.~\ref{sec_par}. An attractive magnetic interaction is induced between the particles due to an external magnetic field. In reality, such an external magnetic field could be applied, for instance, by a conventional permanent magnet. Yet, the magnetic fields generated by such magnets are in general non-homogeneous in space. Due to the induced magnetic moments, the embedded particles are drawn into the magnetic field gradient \cite{jackson1962classical}. Placing the permanent magnet underneath the substrate, symmetrically below the two particles, see Fig.~\ref{fig_reversal}(a), a force arises that pulls the particles approximately perpendicularly towards the surface. 
%
In our geometry, this force is given by 
\begin{equation}
\mathbf{F}_{\nabla} = \mathbf{\hat{z}}\,m\,\partial_zB,
\end{equation}
where $\partial_zB<0$ is the normal gradient in the magnitude of the magnetic field generated by the magnet. 

We plot the relative change in distance $\Delta d/d$ between the particles as a function of their height $h$ above the wall in Fig.~\ref{fig_reversal}(b). Here, we set $\partial_zB=-0.3\sqrt{\mu\mu_0}/a$, $M=4\sqrt{\mu/\mu_0}$, and $d=5a$. First, we recognize that in the bulk, without the influence of the wall, $\Delta d/d<0$. Thus the particles attract each other as expected. However, close to the wall, the situation can be inverted. There, we find values of $h$ for which $\Delta d/d>0$. Thus the particles appear to repel each other. This effective conversion in particle interaction is stronger in maximal magnitude for the free-slip condition and for an incompressible elastic matrix, i.e., for $\nu=0.5$. 

Inspection of Eqs.~(\ref{B})--(\ref{Green_fs}) and (\ref{Mnsii})--(\ref{Mfsij}) reveals that, for each particle, it is mainly the normal mirror-image force of the other particle that causes the effective outward displacement. For both boundary conditions, free-slip and no-slip, the normal gradient forces are mirrored. Thus, for each image particle, the normal force points upwards towards the respective real particle, see Fig.~\ref{fig_reversal}(a). The displacement field induced by this normal mirror force pushes the real particle away from the wall. Simultaneously, however, it pushes the other real particle outwards. 

Illustratively, the effect is readily understood in the following way. Through the gradient force, the particles are pushed against the boundary. This squeezes the elastic matrix between the particles and the wall. Due to its limited compressibility, the matrix material needs to escape to the sides. It partially takes the embedded particles along during this outward displacement. 

We should add a discussion concerning the magnitude of the relative changes in distance $\Delta d/d$ in Fig.~\ref{fig_reversal}(b). The effect of inversion is obvious, but the magnitudes are rather small. This choice was on purpose. Using instead the same parameters as in Fig.~\ref{fig_parallel} together with $\partial_zB=-0.9\sqrt{\mu\mu_0}/a$, we obtain relative changes in distance of the same magnitude as in Fig.~\ref{fig_parallel}, still with the attraction effectively converted into repulsion. However, the normal displacements perpendicular to the wall then become larger than our linearly elastic treatment would safely allow for. Therefore, we here deliberately reduced the magnitudes in the plotted results. 

In SI units, our chosen parameter values correspond, for instance, to a particle radius of $a=100~\mu\mathrm{m}$ and $\partial_zB=-33~\mathrm{T}/\mathrm{m}$. A value of $\partial_zB=-0.9\sqrt{\mu\mu_0}/a$ then corresponds to $-100~\mathrm{T}/\mathrm{m}$. The latter appears relatively large, yet simple estimates yield field gradients on the order of this magnitude on the surfaces of commercially available permanent magnets that have been used in recent experiments \cite{huang2016buckling,puljiz2016forces}. 
Moreover, it is particularly the lowest-order term given by Eq.~(\ref{M0}) that the gradient force competes with concerning its inverting effect. Therefore, increasing the particle radius and simultaneously decreasing the field gradient, e.g., to values $a=1~\mathrm{mm}$ and $\partial_zB=-3.3~\mathrm{T}/\mathrm{m}$, leads to approximately the same results. 
Altogether, we are confident that the effect can be observed in corresponding future experiments. 

A similar overall phenomenon of effective inversion of interactions by a rigid wall had been observed in low-Reynolds-number hydrodynamics in terms of ``like-charge attraction'' \cite{squires2000like}. There, equally charged colloidal particles in parallel configuration close to a rigid no-slip surface were observed to effectively attract each other, despite their mutual electrostatic repulsion. In that situation, the underlying cause were equivalent charges on the wall. By these charges, the particles were pushed away from the surface. Hydrodynamic interactions between the rigid substrate and the particles, mediated by the suspending fluid, then lead to an apparent attraction between the actually repulsive particles.

\section{Conclusions}
\label{conclusion}

In summary, we have considered the situation of particle-laden elastic media in the close vicinity of a rigid wall. Our main focus was on no-slip and free-slip surface conditions, but we also compared to the bulk behavior. 
In a first step, we have transferred the low-Reynolds-number  hydrodynamic approach by Blake \cite{blake1971note} to the linearly elastic case. That is, the Green's function in the presence of a no-slip boundary has been derived by direct calculation via in-plane Fourier transforms. For this purpose, an interim auxiliary variable was introduced in analogy to the hydrodynamic pressure field. 
Our approach adds to further connecting these two related subfields of classical continuum mechanics with each other. 

After illustrating the resulting mirror-image systems beyond the wall, we described the coupled displacements of the embedded particles taking into account their matrix-mediated interactions. In particular, we analyzed the influence of no-slip and free-slip rigid walls on the displacement of two magnetically interacting embedded particles. We considered their mutual arrangement to be parallel and normal to the boundary. From this, we concluded that free-slip substrate conditions generally support the particle relocation and should therefore be preferred in several applications, for example soft actuators \cite{an2003actuating,filipcsei2007magnetic, fuhrer2009crosslinking,bose2012soft}. Finally, we demonstrated that interactions with the wall can even qualitatively affect the particle behavior. For instance, it can convert induced attraction between the particles into effective repulsion. As a motivation, we here considered magnetic particle interactions. However, the approach is more general and in principle any, not necessarily pairwise type of forces on the inclusions can be addressed in the same way. 

Generally, our results are readily verifiable by corresponding experiments. For instance, magnetizable particles could be placed within a soft elastic gel matrix \cite{puljiz2016forces} 
close to a rigid surface. No-slip boundary conditions are frequently satisfied automatically due to strong adsorption interactions with the substrate. In contrast to that, free-slip conditions could be achieved by appropriate lubrication of the surface. Mutual forces between the particles can be induced by external magnetic fields \cite{puljiz2016forces}. 
The displacements of magnetic particles of colloidal size \cite{klapp2016collective} could be tracked for instance by confocal microscopy \cite{huang2016buckling,huang2016microgels} or by x-ray microtomography \cite{gunther2011x,gundermann2014investigation,gundermann2017}. Large enough particles could be directly observed by optical microscopy. 
An additional attraction towards the surface, see Sec.~\ref{reversal}, could be realized by a gradient in the external magnetic field. 

In the future, our approach opens the way to transfer further related solution methods from the hydrodynamic to the linearly elastic case. For instance, the situation of thin elasto-magnetic membranes could be considered \cite{bohlius2008rosensweig,raikher2008shape}. Since linear elasticity theory is a general symmetry-based continuum description, no specific chemical properties of the material need to be satisfied for our approach to apply. 
Thus, our results may even be helpful for the characterization of biological cells and biological tissue on a substrate, where in parts related theoretical strategies have been applied \cite{bischofs2003cell,bischofs2004elastic}. There, the forces can be induced actively by the cells themselves, typically modeled by force dipoles \cite{schwarz2002elastic,schwarz2013physics}, and do not necessarily need to be imposed from outside. For instance, this happens by construction during cell migration \cite{du2005force,tanimoto2014simple}. Large self-organized assemblies of biological cells on rigid substrates are given by biofilms that are usually reinforced by an embedding extracellular polymeric matrix \cite{smalyukh2008elasticity,flemming2010biofilm}. 
Activity-induced elastic interactions between individual cells can influence their overall coordination \cite{yuval2013dynamics} and can be described using the Green's function technique as well \cite{bischofs2004elastic}. 

From an engineering application point of view, our message is the following. Soft magneto- or electrostrictive actuation devices based on the considered principles should in many cases be realized with free-slip boundary conditions, possibly by appropriate lubrication of the substrate. Then it should be possible to achieve larger amplitudes of deformation.

\begin{acknowledgments}
The$\,$ author$\,$ thanks$\,$ the$\,$ Deutsche$\,$ Forschungsgemein- 

\noindent schaft for support of this work through the priority program SPP 1681 (No.\ ME 3571/3).
\end{acknowledgments}


%
%
%

\end{document}